\documentclass[11pt,a4paper]{elsarticle}
\let\proof\relax
\let\endproof\relax
\usepackage{algorithm}
\usepackage{algorithmic}
\usepackage{amsthm}
\usepackage{amsmath}
\usepackage{amssymb}
\usepackage{amsfonts}
\usepackage{mathrsfs}
\newtheorem{thm}{Theorem}

\newtheorem{defn}[thm]{Definition}

\newtheorem{prop}[thm]{Proposition}

\newtheorem{cor}[thm]{Corollary}

\theoremstyle{remark}

\newtheorem{rem}{Remark}

\journal{}

\begin{document}
\begin{frontmatter}

\title{A Multilevel Approach for the  
Performance Analysis of Parallel Algorithms}
\author[label2]{L. D'Amore\corref{cor1}}
\author[label2]{V. Mele}
\author[label3]{D. Romano}
\author[label2]{G. Laccetti}
\address[label2]{University of Naples, Federico II, Naples (IT)}
\address[label3]{Institute of High Performance Computing and Networking (ICAR), CNR, Naples (IT)}

%% \ead[url]{home page}
\cortext[cor1]{Corresponding Author}

\renewcommand{\thefootnote}{\fnsymbol{footnote}}

%\footnotetext[5]{Support in common for the first and second authors.}

\renewcommand{\thefootnote}{\arabic{footnote}}

\setlength{\parindent}{0in}

\begin{abstract}
We provide a multilevel approach for analysing performances of parallel
algorithms. The main outcome of such approach is that the algorithm is described
by using a set of operators which are related to each other according to the problem
decomposition. Decomposition level determines the granularity of the algorithm. A
set of block matrices (decomposition and execution) highlights fundamental characteristics
of the algorithm, such as inherent parallelism and sources of overheads.
\end{abstract}

\end{frontmatter}
\thispagestyle{plain}
\markboth{L.D'Amore et al.}{  }

\section{Introduction and Motivation}

  Numerical algorithms are at the heart of the software that enable scientific discoveries. The development of effective  algorithms has a tremendous impact on harnessing emerging computer architectures to achieve new science.
%Advanced computer architectures are centered around the concept of parallel processing.
The mapping problem, first considered in 1980s \cite{bokari81}, refers to the implementation of algorithms on a given target architecture which is capable to  maximize some performance metrics \cite{berman87,berman88,aggarwal87,miller14,mdhpcl2013}.
 Due to the multidimensional heterogeneity of modern architectures, it is becoming increasingly clear that using the performance metrics  in a one-size-fits-all approach fails to discover sources of performance degradation that hamper to deliver the desired performance level. We believe that a performance model based on  problem-specific features, as well as on  mathematical tools to better analyze and understand  algorithm behavior, should be developed.  The present  article attempts to collect our efforts in this area.\\
 We briefly summarize how the performance model we provide in this work originates. We firstly address the basic structural features of  algorithms which are  dictated by  data and operator  dependencies \cite{flynnInd}. These dependencies refer to relations among computations which need to be satisfied in order to compute the problem solution  correctly. The absence of dependencies indicates the possibility of parallel computations. So the study of data dependencies in an algorithm becomes the most critical step in parallelising the computations of the algorithm. Then, in analogy to the graph of dependency between tasks, we introduce the algorithm as a set of operators starting from a predetermined decomposition of the problem described by a suitably defined matrix, called \emph{decomposition matrix}.  The mapping of the algorithm on the computing machine is described by the \emph{execution matrix}.

\subsection{Organization of the article}
Section \ref{secPrel} will review basic concepts and definitions useful for setting up the mathematical framework. We define  the decomposition matrix; following \cite{flynnInd}, we describe a  parallel algorithm as an ordered set of operators,  moreover we give the  definition of complexity of the algorithm depending on the number of such operators; finally,  we define the execution matrix describing the mapping of the algorithm on the target computing resource. Section \ref{secMet} focuses on two metrics characterizing the algorithm performance, such as the  scale up factor and the  speed up.  In Section \ref{secPer} we analyse the performance of parallel algorithms arising from the same problem decomposition. We derive the Generalized Amdhal's Law and some important upper and lower bounds of the performance metrics. In Section \ref{secPer2} we consider the particular case where the operators of an algorithm  have the same execution time (namely, the operators are the usual floating point operations); in other words, we are assuming to get a decomposition at the lowest level of granularity and we derive the standard expressions for the performance metrics.   In Section \ref{secConc} conclusions are drawn.

\subsection{Related works}
The appropriate mapping depends upon both the specification of the algorithm and the underlying architecture. Firstly, it implies a transformation of the algorithm into an equivalent but more appropriate form. Works on the mapping problem can be classified according to  the used representation.
Graph based approaches perform transformations on the algorithm and the architecture represented as graphs. In this approach the algorithm is modeled in terms of graphs structures and the mapping in terms of graphs partitions \cite{bokari81}. Linear algebra approaches represent the graph and its data dependencies by a matrix, then transform the graph by performing matrix operations. Language based approaches transform one form of program text into another form, where the target form textually incorporates information about the architecture \cite{kuck81}.  Characteristic based approaches represent the algorithm in terms of a set of characteristics which determines the transformations. Included in this category is the work of \cite{Moldovan}, where a technique which abstracts a computation in terms of its data dependencies is described. The method is based on a mathematical transformation of the index sets and of the data-dependency vectors associated with the given algorithm.

One common issue of the aforementioned approaches is that very often the model used for the representation of the algorithm cannot be explicitly employed for deriving the expression of the algorithm's performance metrics. On the contrary, performance analysis is often accomplished with automatic tools on a combination of the algorithm and the parallel architecture on which it is implemented (the so-called parallel system), exploiting automating mappings, automatic translations, re-targeting mappings tracing, auto-tuning tools (such as: the PaRSEC runtime system \cite{keyPARSEC}, that  provides a portable way to automatically adapt algorithms to new hardware trend.  Nevertheless, these approaches ignore the properties of the problem decomposition. Instead, our  model allows to choose a level of abstraction of the problem decomposition and of the algorithm description which determine the level of granularity of the performance analysis.  A set of parameters are used both to describe the problem and to compute speed up, efficiency, cost, overhead, scale up  and  operating point of the algorithm, starting from the problem decomposition. Metrics and their asymptotic estimates,  which represent upper or lower bounds of the  algorithm's performance  depend on  parameters characterizing the structure of the two matrices, namely their number of rows and columns, and on computing environment parameters, such as the execution time for one floating point operation.

\section{Preliminary Concepts and Definitions}\label{secPrel}
We  introduce a dependency relationship among component parts of a computational problem, among  operators of the algorithm that solves the problem and, finally, among memory accesses of the algorithm. In this way we are able to define two matrices (decomposition and execution) which  highlights fundamental characteristics of the algorithm and  which are the foundations of the mathematical model we are going to introduce.
To this aim we first give  some definitions which we refer in this work\footnote{It is worth to note that these definitions do not claim to be general. Their aim is to establish the mathematical setting on which we will restrict our attention.}.

 \begin{defn}(\textbf{Computational Problem})
A computational problem $\mathcal{B}_{N_r}$ is the mathematical problem specified by an input/output function:
$$\mathcal{B}_{N_r}:\mathit{In}_{\mathcal{B}_{N_r}}\mapsto \mathit{Out}_{\mathcal{B}_{N_r}},$$
 where $N_r$ is the input data size and $r\in \mathbb{N}$, between the data of and the solution of $\mathcal{B}_{N_r}$.
\end{defn}

Therefore, in the following  we assume that the computational problem  $\mathcal{B}_{N_r}$ is identified by the triple:
$$\mathcal{B}_{N_r}\equiv\left(N_r,\mathit{In}_{\mathcal{B}_{N_r}},\mathit{Out}_{\mathcal{B}_{N_r}}\right)$$

\begin{defn}(\textbf{Similar Computational Problem})\label{similar}
Two computational problems, $\mathcal{B}_{N_r}$ and $\mathcal{B}_{N_q}$, are said similar if they are specified by the same functional relation and they only differ in the input/output data size. If $\mathcal{B}_{N_r}$ and $\mathcal{B}_{N_q}$ are similar we write $\mathcal{B}_{N_r}\mathscr{S} \mathcal{B}_{N_q}$.
\end{defn}

Dividing a computation into smaller computations, some or all of which may potentially be executed in parallel, is the key step in designing parallel algorithms. The parts that a problem is decomposed into often share input, output, or intermediate data. The dependencies usually result from the fact
that the output of one part is the input for another. In our mathematical framework the relationship among component parts of a computational problem will be described by the so called decomposition matrix. In order to define this matrix we  need to introduce the following algebraic structure

\begin{defn}(\textbf{Dependency Group})\label{dipgroup}
Let $(\mathcal{E}, \pi)$ be a group and let  $\pi_{\mathcal{E}}$ be a strict partial order relation on $\mathcal{E}$, which is compatible with $\pi$.
We say that any  element of $\mathcal{E}$, let us say $A$, depends on an element of $\mathcal{E}$, let us say  $B$, if $A \pi_{\mathcal{E}} B$, and we write  $A \leftarrow B$. If $A$ and $B$ do not depend on each other we write $A \nleftarrow B$. The group $(\mathcal{E},\pi)$ equipped with $\pi_{\mathcal{E}}$ is called dependency group and it is denoted as $(\mathcal{E},\pi, \pi_{\mathcal{E}})$.
\end{defn}

%\emph{Remark}:
\begin{rem}
    Since $\pi_{\mathcal{E}}$ is transitive, from Definition \ref{similar} it follows that any  two elements of $\mathcal{E}$, let us say $A$ and $B$,  are independent if there is no any relationship  between them. In this case  we write $A \nleftarrow B$ and $B \nleftarrow A$, or even $A \nleftrightarrow B$.
\end{rem}

\noindent Now we are able to define the dependency matrix on $(\mathcal{E},\pi, \pi_{\mathcal{E}})$.
\noindent   \begin{defn}(\textbf{Dependency Matrix})\label{dipmat}
Given $(\mathcal{E}, \pi, \pi_{\mathcal{E}})$, the matrix\footnote{For simplicity of notation, in the following we will continue to define matrices in the usual sense of matrix calculation; seen as a family, dependency matrix is defined by the triple:
$$\mathcal{F}= ((\mathcal{E}, \pi, \pi_{\mathcal{E}}),[0,r_D-1] \cdot[0,c_D-1], f)$$ where $f$ is an application between $(\mathcal{E},\pi, \pi_{\mathcal{E}})$ and the set of indices.}
$\mathcal{F}$,  of size  $r_D\cdot c_D$, whose elements $d_{i,j} \in (\mathcal{E}, \pi)$,
are such that $\forall i\in[0,r_D-1]$
\begin{equation}\label{matrix}
 d_{i,j} \nleftrightarrow d_{i,s} \quad, \quad   \forall  s,\,j\in[0,c_D-1]\\
\end{equation}
and $\forall i\in[1,r_D-1], \quad  \exists q \in [0,c_D-1] $\, s.t.
\begin{equation}\label{matrixbis}
d_{i,j}\leftarrow d_{i-1,q} \quad , \quad \forall j \in [0, c_D-1],
\end{equation}
while the others elements are set equal to zero, is said the dependency matrix. %Without loss of generality we can consider matrix $\mathcal{F}$ as unique.
\end{defn}

%\noindent \emph{Remark:}
\begin{rem}
    Matrix $\mathcal{F}$ is unique (through its construction), up to a permutation of elements on the same row.
$c_{\mathcal{F}}$ is said the concurrency degree\footnote{A similar concept has already been highlighted in \cite{Gupta}} of  $(\mathcal{E},\pi, \pi_{\mathcal{E}})$ and  $r_{\mathcal{F}}$ is the said the  dependency degree of  $\mathcal{E}$. Concurrency degree measures the intrinsic concurrency among sub-problems of $(\mathcal{E}, \pi, \pi_{\mathcal{E}})$. It is obtained as the number of  columns of $\mathcal{F}$. %Therefore, if there are only non zero elements $(\mathcal{E}, \pi, \pi_{\mathcal{E}})$ has the highest intrinsic concurrency.
\end{rem}

\subsection{The Problem Decomposition}
\noindent Let  $S({\mathcal{B}_{N_r}})$  denote the solution\footnote{Here, for the sake of simplicity,  we assume that
 $S(\mathcal{B}_{N_r})$ exists and it is unique.} of  $\mathcal{B}_{N_r}$.

    \begin{defn}(\textbf{Decomposition of a computational problem})\label{defdec}
Given $\mathcal{B}_{N_r}$, any finite set of computational problems $\{\mathcal{B}_{N_i}\}_{i=0, \ldots,k-1}$, where $ k\in \mathbb{N}$, such that $\mathcal{B}_{N_r} \leftarrow \mathcal{B}_{N_i}$, where $N_i<N_r$,   and
	$$\sum_{i=0}^{k-1}N_i\ge N_r \quad , $$
	 is called  a decomposition of $\mathcal{B}_{N_r}$. $\mathcal{B}_{N_i}$ denotes a sub-problem  of $\mathcal{B}_{N_r}$. A decomposition of $\mathcal{B}_{N_r}$, which  is denoted as
		\begin{equation}\label{decomp1}
			D_{k}(\mathcal{B}_{N_r}):=\{\mathcal{B}_{N_0},\ldots,\mathcal{B}_{N_{k-1}}\},
		\end{equation}
defines the computational problem
$$  	D_{k}(\mathcal{B}_{N_r})\equiv (\sum_{i=0}^{k-1}N_i, In_{\mathcal{B}_{N_r}}, Out_{\mathcal{B}_{N_r}})\quad .$$
 The  set of all the decompositions  of $\mathcal{B}_{N_r}$ is denoted as $\mathcal{D}\mathcal{B}_{N_r}$ .
    \end{defn}

 \begin{defn}(\textbf{Similar Decompositions})\label{simdec}
       Given  $\mathcal{B}_{N_r}\mathscr{S}\mathcal{B}_{N_q}$, two decompositions  $D_{k_i}(\mathcal{B}_{N_r})$ and $D_{k_j}(\mathcal{B}_{N_q})$  are called  similar if

       $$k_i=card(D_{k_i}(\mathcal{B}_{N_r}))=card(D_{k_j}(\mathcal{B}_{N_q}))=k_j$$
       and $$ \forall \,\mathcal{B}_{N_s} \in D_{k_i}(\mathcal{B}_{N_r}) \, \exists \, ! \, \mathcal{B}_{N_t} \in D_{k_j}(\mathcal{B}_{N_q}) \, : \mathcal{B}_{N_s} \mathscr{S} \mathcal{B}_{N_t} \quad ,$$
       and we write $$D_{k_i}(\mathcal{B}_{N_r})\mathscr{S} D_{k_j}(\mathcal{B}_{N_q})\quad . $$

        %$$k_j+1=p\cdot (k_i+1)+q,\quad  q\in \mathbb{N},\quad p\in \mathbb{N},\quad p>1$$
\end{defn}

%\noindent \emph{Remark (\textbf{Decomposition matrix})}:

\begin{rem} \emph{(\textbf{Decomposition matrix})}\label{remdec}
In order to
capture  interactions among  component parts (or sub-problems) of  $\mathcal{B}_{N_r}$, we use the dependency matrix on $D_{k}(\mathcal{B}_{N_r})$.
More precisely, by using Definition 2 we introduce the group   $(D_{k}(\mathcal{B}_{N_r}), g_{sol})$ where $g_{sol}$ is any application  between any two elements $\mathcal{B}_{N_i}$  and $\mathcal{B}_{N_j}$ of $D_{k}(\mathcal{B}_{N_r})$, equipped with the strict partial order relation $\pi_{D_{k}(\mathcal{B}_{N_r})}$. Then, we construct  the (unique) dependency matrix $\mathcal{F}$ corresponding to the decomposition $D_{k}(\mathcal{B}_{N_r})$. In the following we denote this matrix as $M_D(D_{k}(\mathcal{B}_{N_r}))$,  or $M_{D_k}$ for simplicity,  and we refer to it as the \textbf{decomposition matrix}. Given $D_{k}(\mathcal{B}_{N_r})$, let  $c_{D_k}$ denote the number of columns. This is the (unique) concurrency degree of  $\mathcal{B}_{N_r}$. Let $r_{D_k}$ denote the row number of rows. This is the (unique)  dependency degree of  $\mathcal{B}_{N_r}$. Concurrency degree measures the intrinsic concurrency among sub-problems of $\mathcal{B}_{N_r}$. %
\end{rem}

\noindent We observe that, if there are not empty elements, the problem $\mathcal{B}_{N_r}$ has the highest intrinsic concurrency, hence we give the following
    \begin{defn}\textbf{(Perfectly Decomposed Problems)}\label{ProbPar}
$\mathcal{B}_{N_r}$ is said perfectly decomposed if  $\exists D_{k}(\mathcal{B}_{N_r})$ and  $M_D$ such that
    \begin{itemize}
      \item  $c_D>1$ ;
      \item $\forall \, i,j$,  $d_{i,j}\neq \emptyset $.
    \end{itemize}
   \end{defn}

\normalsize

\noindent The next step is to take these parts and assign them (i.e., the mapping step)  onto the computing machine. In the next section we introduce the computing environment characterized by the set of logical-operational operators/operations that it is able to apply/execute.

\subsection{The computing architecture}\label{subsecComp}
We introduce  the machine  $\mathcal{M}_P$ equipped with $P \geq 1$ processing elements with specific logical-operational capabilities such as: basic operations (arithmetic,$\ldots$), special functions evaluations ($\sin,\cos,\ldots$), solvers (integrals, equations system, non linear equations$\ldots$). These are the computing operators of $\mathcal{M}_P$. In particular, we will use the following characterization of operators of $\mathcal{M}_P$.

 \begin{defn}(\textbf{Computing Operators})
	The operator $I^{j}$ of $\mathcal{M}_P$ is a correspondence between $\mathbb{R}^s$ and $\mathbb{R}^t$, where $s,\,t \in \mathbb{N}$ are
 positive integers. %\\[.1cm]
\end{defn}

\noindent Given  $\mathcal{M}_P$, the set without repetitions
	$$Cop_{\mathcal{M}_P}:=\{I^{j}\}_{j\in[0,q-1]},$$
	where  $q\in\mathbb{N}$,  characterizes logical-operational capabilities of the machine $\mathcal{M}_P$.
\noindent   Operators, properly organized, provide the solution to $\mathcal{B}_{N_r}$,  as stated in the following

    \begin{defn}(\textbf{Solvable Problems})\label{risolvibile}
	$\mathcal{B}_{N_r}$ is  solvable in $\mathcal{M}_P$ if
	$$\exists D_{k}(\mathcal{B}_{N_r})\in \mathcal{DB}_{N_r} \,: \forall \mathcal{B}_{N_j}\in D_{k}(\mathcal{B}_{N_r}) \qquad \exists I^{j}\in Cop_{\mathcal{M}_P}\, : I^{j}[\mathcal{B}_{N_j}]=S(\mathcal{B}_{N_j})$$
	that is, if it exists any relation
	\begin{equation}\label{theta}
	\theta: \mathcal{B}_{N_j} \in D_{k}(\mathcal{B}_{N_r})\in \mathcal{DB}_{N_r} \longmapsto I^{j}\in Cop_{\mathcal{M}_P}.
	\end{equation}
	%La decomposizione $D$ \`e detta \emph{decomposizione elementare o atomica}.
    \end{defn}

\noindent In particular, we say that a decomposition is suited for $\mathcal{M}_P$ if $\theta$ is a function. From now on, we consider as solvable any  problem $\mathcal{B}_{N_r}$, and as fixed any decompositions  $D_{k}(\mathcal{B}_{N_r})\in \mathcal{DB}_{N_r}$ suited for $\mathcal{M}_P$. \footnote{Note that there is no loss of generality.}.

 \noindent We associate execution time $t_{i}$ (measured, for instance, in seconds) to each  $I^{i}\in Cop_{\mathcal{M}_P}$
   in $\mathcal{M}_P$. If  $I^{i}\equiv \varnothing$, we set $t_{\varnothing}=0$.

\subsection{The Algorithm }\label{secalg}

\noindent In the literature,  an  algorithm is any procedure consisting of  finite number of unambiguous rules that specify a finite sequence of operations to reach a solution to a problem or a specific class of problems \cite{kronsjo}. Here we define an algorithm as a proper set of operators  which solves  $\mathcal{B}_{N_r}$, as stated in the following
\begin{defn}(\textbf{Algorithm})\label{algo}
Given  $D_{k}(\mathcal{B}_{N_r})$,  an  algorithm  solving  $\mathcal{B}_{N_r}$, indicated as
	$$A_{D_{k}(\mathcal{B}_{N_r}),\mathcal{M}_P}=\{{I^{i_0},I^{i_1},...I^{i_k}\}}$$
is a sequence of elements (not necessarily distinct) of $Cop_{\mathcal{M}_P}$,
	such that \footnote{In the following we use the symbol $\circ$ to denote correspondence composition.}
	$$I^{i_k} \circ I^{i_{k-1}} \circ ... \circ I^{i_0} [\mathcal{B}_{N_r}]= S(\mathcal{B}_{N_r}),$$
	where $j\in[0,card(Cop_{\mathcal{M}_P})-1]$, and such that there is a bijective correspondence
	\begin{equation}\label{gamma}
	\gamma: \mathcal{B}_{N_\nu} \in D_{k}(\mathcal{B}_{N_r})\in \mathcal{DB}_{N_r} \longleftrightarrow I^{i_j}\in A_{D_{k}(\mathcal{B}_{N_r}),\mathcal{M}_P}
	\end{equation}
    Every ordered subset of  $A_{D_{k}(\mathcal{B}_{N_r}),\mathcal{M}_P}$ is a sub-algorithm of $A_{D_{k}(\mathcal{B}_{N_r}),\mathcal{M}_P}$.
    \end{defn}

 \noindent For simplicity of notations and when there is no  ambiguity, we indicate  algorithms briefly as $A_{k,P}$.

    \begin{defn}(\textbf{Equal Algorithms)}\label{alguguali}
    Two algorithms
     $$A_{k,P}^i=\{{I^{i_0},I^{i_1},...I^{i_k}\}}, \quad A_{k,P}^j=\{{I^{j_0},I^{j_1},...I^{j_k}\}}$$
     are said equal if $\forall s\in[0,k] , \quad I^{i_s}\equiv I^{j_s}.$
      \end{defn}
% \end{minipage}}
%\vspace{0.1cm}
\noindent Note that two equal algorithms have the same cardinality.\\

    \begin{defn}(\textbf{Granularity set of an Algorithm)}\label{alguguali}
    Given $A_{k,P}$, the subset $\mathcal{G}(A_{k,P})$ of  $A_{k,P}$ made of distinct operators of  $A_{k,P}$ defines the granularity set of $A_{k,P}$.
    Two algorithms
     $$A_{k,P}^i=\{{I^{i_0},I^{i_1},...I^{i_k}\}}, \quad A_{k,P}^j=\{{I^{j_0},I^{j_1},...I^{j_k}\}}$$
     have the same granularity  if $\mathcal{G}(A_{k,P}^i)\equiv \mathcal{G}(A_{k,P}^j)$.
      \end{defn}
% \end{minipage}}

\noindent  Let  $AL_{\mathcal{B}_{N_r}}$ (or simply $AL$) be the set  of  algorithms that solve $\mathcal{B}_{N_r}$, obtained by varying  $\mathcal{M}_P$, the number of processing units $P$ and  $D_{k}(\mathcal{B}_{N_r})\in \mathcal{DB}_{N_r}$. Even if one can easily formulate infinite variations of an algorithm that do the same thing, in the following we assume $AL$ to be finite.\\

\begin{defn}(\textbf{The quotient set $\frac{AL}{\varrho}$})
%Let associate each algorithm of $AL$ to a decomposition suited for $\mathcal{M}_P$, that is,
Let
\begin{equation}\label{phi}
\varphi: A_{k,P} \in AL\longrightarrow D_{k}(\mathcal{B}_{N_r})\in \mathcal{DB}_{N_r},
\end{equation}
be the surjective correspondence which  induces on  $AL$ an equivalence relationship $\varrho$ of  $AL$ in itself, such that
\begin{eqnarray}
      \varrho(A_{k,P})= \{\widetilde{A_{k,P}}\in AL : \varphi(\widetilde{A_{k,P}})&=&\varphi(A_{k,P})\}.
       \end{eqnarray}
 The set $\varrho(A_{k,P})$ consists of algorithms of  $AL$ associated  with the same decomposition $D_{k}(\mathcal{B}_{N_r})\in \mathcal{DB}_{N_r}$.
 $\varrho$ induces  the quotient set $\frac{AL}{\varrho}$, whose elements are disjoints and finite subsets of $AL$ determined by  $\varrho$, that is they are  equivalence classes under  $\varrho$.\\
\end{defn}
%\vspace{0.1cm}

\noindent In the following we assume $A_{k,P}$ to represent its equivalence class in $AL$.

    \begin{defn}(\textbf{Complexity})  \label{Compl}	
The cardinality of $A_{k,P}$, denoted as $C(A_{k,P})$,  is said complexity of $A_{k,P}$. It is
$$C(A_{k,P}) :=card(A_{k,P})=k\quad .$$
    \end{defn}
%\end{minipage}}

%\vspace{0.1cm}
%\noindent \emph{Remark}:

\begin{rem}
    $C(A_{k,P})=k$ equals to the number of non empty elements of $M_{D_k}$, i.e. the decomposition matrix defined on $D_{k}(\mathcal{B}_{N_r})$. By virtue of the bijective correspondence $\gamma$ in (\ref{gamma}), it holds that
    \begin{equation}\label{Card1}
    card(A_{k,P})=card(D_{k}(\mathcal{B}_{N_r}))=k, \quad \forall \, A_{k,P}\in \varrho(A_{k,P})\,.
    \end{equation}
    So,  each algorithm belonging to the same equivalence class according to $\varrho$ has the same complexity. An integer (the complexity) is therefore associated with each element $\varrho(A_{k,P})$ of quotient set $\frac{AL}{\varrho}$ which induces an ordering relation between the equivalence classes in  $\frac{AL}{\varrho}$: therefore there is a minimum complexity for algorithms that solve the problem $\mathcal{B}_{N_r}$.
\end{rem}

%\noindent \emph{Remark} (\textbf{Similar Algorithms}):
\begin{rem}(\textbf{Similar Algorithms}) Given $\mathcal{B}_{N_r} \mathscr{S} \mathcal{B}_{N_q}$ and their relative similar decompositions
    $D'_{k_i}(\mathcal{B}_{N_r})\mathscr{S} D^{''}_{k_j}(\mathcal{B}_{N_q})$ with $k_i,k_j\in \mathbb{N}$ and $k_i=k_j=k$ (see Definition \ref{simdec}), algorithms belonging to
    $\varrho(A_{k_i,P})=\varphi^{-1}(D'_{k_i}(\mathcal{B}_{N_r}))$
    (see (\ref{phi})) are  similar to algorithms belonging to
    $\varrho(A_{k_j,P})=\varphi^{-1}(D^{''}_{k_j}(\mathcal{B}_{N_q}))$.
    From Definition \ref{simdec} and \ref{Compl} and the (\ref{Card1}), it follows that
    $$A_{k_i,P}\mathscr{S} A_{k_j,P} \Longrightarrow C(A_{k_i,P})=C(A_{k_j,P})=k$$
    that is similar algorithms have the same complexity.
\end{rem}

%\noindent \emph{Remark}:
\begin{rem}
     As we can associate   $I^{i_k}\in A_{k,P}$ to each subproblem according to  $\gamma$, then the operators of $A_{k,P}$ inherit the dependencies existing between subproblems of $\mathcal{B}_{N_r}$, but they do not inherit independencies, because for instance,  two operators may depend on the availability of computing units of $\mathcal{M}_P$ during their execution \cite{flynnInd}.
\end{rem}

%\noindent \emph{Remark (\textbf{Execution matrix})}:

\begin{rem} (\textbf{Execution matrix}) \label{exmat}
According to Definition \ref{dipgroup}, we introduce the group $\left(\mathcal{P}\left(A_{k,P}\right), \circ, \pi_{A_{k,P}}\right)$ where $\mathcal{P}\left(A_{k,P}\right)$ is the set of all the sub-algorithms of $A_{k,P}$, and $\pi_{A_{k,P}}$ is the strict partial order  relation  between any two elements of $\mathcal{P}\left(A_{k,P}\right)$ that guarantees that two elements  cannot be  performed in any arbitrary order and simultaneously\footnote{The  condition that two elements cannot be  performed in any arbitrary order induces the inheritance of dependencies between decomposition subproblems and algorithm operators, while the condition that two elements cannot be  performed simultaneously - relating to  availability of resources - adds possible reasons for dependency between operators, which depend on the machine on which  algorithm $A$ is intended to run \cite{flynnInd}.}.
We construct  matrix $\mathcal{F}$ of order  $r_E\cdot c_E$, where $c_E=P$ \footnote{In general $c_E\leq P$, but we can exclude cases where  dependencies existing between subproblems do not allow to use all the computing units available, i.e. in which  $c_E<P$, because they can easily taken
back to the case where $c_E=P$.} as a dependency matrix (see Definition \ref{dipmat}). The number of columns of this matrix will represent the maximum number of sub-algorithms that can be performed simultaneously on $\mathcal{M}_P$. In the following, we denote this matrix as \textbf{execution matrix} and we refer to it by using the symbol $M_{E}(A_{k,P})=(e_{i,j})$ or simply $M_{E_{k,P}}$ if there is no ambiguity. Matrix $M_{E_{k,P}}$ is unique up to a permutation of elements on the same row. This matrix can be placed in  analogy with the execution graphs (see \cite{keyBern,keyDague,keyCoff,keySharp}) that are often used to describe the sequence of steps of an algorithm  on a given machine for a particular input or a particular configuration.
\end{rem}

\begin{rem}\label{remdimensions}
      As it is $card(A_{k,P})=card(D_{k}(\mathcal{B}_{N_r}))$,  then $M_{D_k}$ and $M_{E_{k,P}}$ have the same number of non empty elements ($k$), whichever is   $P\geq 1$. If $c_E=P=c_{D_k}$, it exists $A_{k,P}$ whose matrix $M_{E_{k,P}}$  has exactly the same structure of the matrix $M_{D_k}$.
\end{rem}

\begin{defn}\label{PurPar}
	 $A_{k,P}$ is said  perfectly parallel if:
	\begin{itemize}
	 \item $c_E>1$;
	 \item $\forall \, i,j \,\,\, e_{i,j} \neq \emptyset $.
	\end{itemize}
	$A_{k,P}$ is said sequential  if:
	\begin{itemize}
	 \item $c_E=1$;
	 \item $\nexists \, j>1\,: \, e_{i,j} \neq \emptyset$.
	\end{itemize}
	$A_{k,P}$ is said (simply) parallel if:
	\begin{itemize}
	 \item $c_E>1$;
	 \item $ \exists \, i,j\,: \, e_{i,j} = \emptyset$.
	\end{itemize}
Moreover,
\begin{itemize}
       \item Every row of  matrix $M_{E_{k,P}}$ such that $\exists \,e_{i,j} \neq \emptyset$, where $j>1$, is a parallel sub-algorithm of $A_{k,P}$.
       \item Every row of  matrix $M_{E_{k,P}}$ such that $\exists \,!\, e_{i,j} \neq \emptyset $  is a sequential sub-algorithm of $A_{k,P}$.
      \end{itemize}
    \end{defn}

%\fbox{\begin{minipage}{\textwidth}
\noindent\begin{rem} Observe that the concurrency degree of  $\mathcal{B}_{N_r}$ in a given decomposition provides an upper limit to the maximum number of independent sub-algorithms executable simultaneously on the machine. The dependency degree provides a lower limit to the execution time of the algorithm.
\end{rem}

\noindent Finally, from  correspondence $\gamma$ (see (\ref{gamma})), we say that  $\mathcal{B}_{N_r}$ is
 solvable in $\mathcal{M}_P \Leftrightarrow \exists \,D_{k}(\mathcal{B}_{N_r})\in \mathcal{DB}_{N_r}\; : \,  \exists\, A_{k,P}$ that solves $\mathcal{B}_{N_r}$ .\\

%\vspace{0.1cm} \fbox{\begin{minipage}{\textwidth}
    \begin{thm}\label{thmProbPar1}
 If  $\mathcal{B}_{N_r}$ is perfectly decomposed according to  $D_k$, $\exists \,\mathcal{M}_P$, where $P>1$, such that $\exists A_{k,P}$  perfectly parallel that solves  $\mathcal{B}_{N_r}$.
      \begin{proof}
      If  $\mathcal{B}_{N_r}$ is perfectly decomposed  then  the matrix $M_{D_k}$ has not empty elements and has order greater than $1$.
      Since  $card(A_{k,P})=card(D_{k}(\mathcal{B}_{N_r}))=k$, it exists $A_{k,P}$ with execution matrix $M_{E_{k,P}}$ of order $r_E \cdot c_E$,  with only non zero elements, such that
      $$r_E =r_{D_k} \text{ and } c_E=P=c_{D_k}$$
      or\footnote{If the concurrency degree $c_{D_k}$ is so great that we can not imagine a real machine with so many units, we can always use a  number of computing units $P=c_{D_k}/n$ with $c_{D_k}\bmod(n)= 0$. This will mean that the  execution matrix of $A_{k,P}$  will have $n$ times more rows and $n$ times less columns than the dependency matrix.}
      $$r_E =n\cdot r_{D_k} \text{ and } c_E=P=c_{D_k}/n$$
      with the integer $n$ is such that $n<c_{D_k}$ and $c_{D_k}\bmod n= 0$.

      In conclusion,
      \begin{itemize}
       \item  $M_{E_{k,P}}$ has $c_E=P>1$ columns,
       \item no rows have an empty element;
      \end{itemize}
      so $A_{k,P}$ is perfectly parallel.
      \end{proof}
    \end{thm}
%\end{minipage}}

%\noindent \emph{Remark}:

\section{Algorithm Performance Metrics}\label{secMet}
In this section we employ the mathematical settings we  introduced in section \ref{secPrel}, in order to define two  quantities to measure the performance of an algorithm: the scale up and the speed up.

\subsection{Scale Up}
Let us consider two decompositions $D_{k_i}(\mathcal{B}_{N})$ and  $D_{k_j}(\mathcal{B}_{N})$ in $\mathcal{DB}_{N}$. Let us consider $A_{k_i,P}$ and $A_{k_j,P}$ representing their equivalence class in $AL$. In order to measure the scalability of parallel algorithms we introduce the following quantity

\begin{defn}(\textbf{Scale up factor})\label{scaleup}
 If $A_{k_i,P}$ and $A_{k_j,P}$  have the same granularity set (see Definition \ref{alguguali}),  the ratio
 	\begin{equation}\label{factor}
 		Sc_{up}(A_{k_i,P},A_{k_j,P}):=\frac{k_i}{k_j}
 	\end{equation} 	
 is said  scale up factor of $\varrho(A_{k_j,P})$ measured  with respect to $\varrho(A_{k_i,P})$.\\
 From Definition  \ref{Compl}, it follows that
 \begin{equation}\label{eq14}
 		Sc_{up}(A_{k_i,P},A_{k_j,P})=\frac{C(A_{k_i,P})}{C(A_{k_j,P})}
    \end{equation}
 \end{defn}

Next proposition quantifies the scale up when we solve the same problem with an algorithm that is the concatenation of several algorithms which are similar to the first one, with polynomial complexity of degree $d$.  \\
\begin{prop}\label{thmScaleUp}
 	Given $\mathcal{B}_{N_r}$, $D_{k}(\mathcal{B}_{N_r})$ and $D_{k'}(\mathcal{B}_{N_r})=\{D_{k'_i}(\mathcal{B}_{N_q})\}_{i=1,\mu}$ where
	\begin{itemize}
	    \item $N_q=N_r/\mu$ with $\mu\in N$, $\mu\leq N_r$, and $N_r\bmod \mu = 0$,
	    \item $\mathcal{B}_{N_q}\mathscr{S} \mathcal{B}_{N_r}$,
	    \item $D_{k}\mathscr{S} D_{k'_i}\mathscr{S} D_{k'_j}$, $\forall i\neq j$.
	\end{itemize}
	Consider $A_{k,P}\in \varphi^{-1}(D_{k}(\mathcal{B}_{N_r}))$ and $A_{k'_i,P}\in \varphi^{-1}(D_{k'_i}(\mathcal{B}_{N_q}))$ and assume that
	\begin{itemize}
	    \item $C(A_{k,P})=k=\mathcal{P}^{d}(N_r)$
	    \item $C(A_{k'_i,P})=k'_i=\mathcal{P}^{d}(N_q)$
	\end{itemize}
	where
 	$$\mathcal{P}^{d}(x)=a_{d} x^{d} + a_{d-1} x^{d-1}+ \ldots + a_0, \quad a_d \neq 0\in \Pi_d\,, x \in \Re$$
	then
	$$Sc_{up}(A_{k,P},A_{k',P})=\xi(N_r,\mu)\cdot \mu^{d-1}$$
	where
	\begin{equation}\label{alpha}
	\xi(N_r,\mu):=\frac{a_{d} + \frac{a_{d-1}}{N_r}+ \ldots + \frac{a_0}{N_r^{d}}}{a_{d} +a_{d-1}\frac{\mu}{N_r}+ \ldots + a_0\frac{\mu^d}{N_r^d}}
	\end{equation}

	\begin{proof} We have that
	\begin{equation}
	    C(A_{k',P})=\sum_{i=1}^\mu C(A_{k'_i,P})= \mu \cdot \mathcal{P}^{d}(N_q)
	\end{equation}
	then from the (\ref{eq14}), it follows that
	\begin{equation}
	    Sc_{up}(A_{k,P},A_{k',P})=\frac{C(A_{k,P})}{C(A_{k',P})}=\frac{\mathcal{P}^{d}(N_r)}{\mu \cdot \mathcal{P}^{d}(N_q)}
	\end{equation}
	that is
	\begin{equation}
	    Sc_{up}(A_{k,P},A_{k',P})=\frac{a_{d} N_r^{d} + a_{d-1} N_r^{d-1}+ \ldots + a_0}{\mu \cdot \left( a_{d} N_q^{d} + a_{d-1} N_q^{d-1}+ \ldots + a_0 \right)}\quad .
	\end{equation}
	Since $N_q=N_r/\mu$, then it is
	\begin{equation}
	    Sc_{up}(A_{k,P},A_{k',P})=\frac{a_{d} (\mu N_q)^{d} + a_{d-1} (\mu N_q)^{d-1}+ \ldots + a_0}{\mu \cdot \left( a_{d} N_q^{d} + a_{d-1} N_q^{d-1}+ \ldots + a_0 \right)}\quad ,
	\end{equation}
	then thesis follows from the (\ref{alpha}).
 \end{proof}
 \end{prop}	

 \begin{cor}
     If $N_r$ is fixed, and $\mu\simeq  N_r$ it is
     $ \xi(N_r,\mu)=const \quad , \,const\in (0,1]$,
     and
      $Sc_{up}(A_{k,P},A_{k',P})\leq N_r^{d-1}$.
     If $\mu$  is fixed, it is
     $$ \lim_{N_r\rightarrow \infty} \xi(N_r,\mu)=const \quad ,\, const\in (0,1]\quad ,$$
     and
      $$ \lim_{N_r\rightarrow \infty}Sc_{up}(A_{k,P},A_{k',P})\leq \mu^{d-1}\quad .$$
     If  $a_i=0,$ $\forall i<d$ then $\xi(N_r,\mu)=1$ and $S_{up}(A_{k,P},A_{k',P})= \mu^{d-1}$, $\forall \mu \quad .$
 \end{cor}

\subsection{Speed Up}
\noindent Let $tcalc$ be the execution time of one  floating point operation.
\begin{rem}
   In the following when we need to refer to execution time of the computing operators of $A_{k,P}$ we will use the following notation of the parameters $\beta^{calc}_{\ldots,M_{E_{k,P}}}$ highlighting the execution matrix $M_{E_{k,P}}$ characterizing the mapping of the algorithm on the machine $\mathcal{M}_P$.
\end{rem}

We assume that
\begin{equation}\label{assunzione}
  \forall I^{i_j}\in Cop_{\mathcal{M}_P}, \quad t_{i_j}=\beta^{calc}_{i_j,M_{E_{k,1}}} \cdot tcalc, \quad \beta_{i_j,M_{E_{k,1}}}^{calc} \in \Re, \quad \beta_{i_j,M_{E_{k,1}}}^{calc}\geq 1
\end{equation}

\begin{defn}(\textbf{Row execution time})\label{TempoR}
	The quantity
	\begin{equation}\label{rowtime}
 T_r(A_{k,P}):=\max_{j\in [0,c_E-1]} t_{r_j}
\end{equation}
is said execution time of the row $r$ of $M_{E_{k,P}}$ (which is a sub-algorithm of $A_{k,P}$). 	
   \end{defn}
%\end{minipage}}

\begin{rem}\label{TempoRrem}
   Let $\beta_{r,M_{E_{k,P}}}^{calc}:= \max_{j\in [0,c_E-1]} \beta^{calc}_{r_j,M_{E_{k,P}}}$ then
$$ T_r(A_{k,P})=\max_{j\in [0,c_E-1]} \beta^{calc}_{r_j,M_{E_{k,P}}} \cdot tcalc = \beta_{r,M_{E_{k,P}}}^{calc} \cdot tcalc \quad .$$
Note that  $\beta_{i_j,M_{E_{k,1}}}^{calc}\geq 1$ then $\beta_{r,M_{E_{k,1}}}^{calc}\geq 1$.
\end{rem}

%\fbox{\begin{minipage}{\textwidth}
    %\begin{defn}[\textbf{Tempo d'esecuzione dell'algoritmo\footnote{In letteratura molti si riferiscono
    \begin{defn}(\textbf{Execution time})\label{TempoE}
	The quantity
	\begin{equation}\label{tesec1}
	 T(A_{k,P}):=\sum_{r=0}^{r_E-1} T_r(A_{k,P})
	\end{equation}
is said   execution time of $A_{k,P}$.
    \end{defn}

\begin{rem}
Let $\beta_{M_{E_{k,P}}}^{calc}:=\sum_{r=0}^{r_E-1}  \beta^{calc}_{r,M_{E_{k,P}}}$  then $\beta_{M_{E_{k,P}}}^{calc}\geq r_E$.
\begin{equation}\label{tesec1bis}
 T(A_{k,P}) = \beta_{M_{E_{k,P}}}^{calc}\cdot tcalc\quad .
\end{equation}
\end{rem}

\begin{rem}\label{alphaall}
    Let
    \begin{equation}\label{betacalsum}
\beta_{sum,M_{E_{k,P}}}^{calc}:=\sum_{i=0}^{r_E-1} \sum_{j=0}^{c_E-1} \beta^{calc}_{i_j,M_{E_{k,P}}}\quad .
\end{equation}

    Then, if $P=1$ then $\beta_{M_{E_{k,P}}}^{calc}:=\beta_{sum,M_{E_{k,P}}}^{calc}$.
\end{rem}

\begin{rem} \label{par+seq}
Let
\begin{itemize}
 \item $r_{seq}\leq r_E$ denote the number of rows of $M_{E_{k,P}}$ with only one non-empty element (sequential sub-algorithms of  $A_{k,P}$).
      %To each row corresponds a row time $T_{i_j}$ with $j\in [0,r_{seq}-1]$;
 \item $r_{par}=r_E-r_{seq}$, with $r_{par}\leq r_E$, denote the number of rows of  $M_{E_{k,P}}$ with more than one non empty element.
     %To each row corresponds a row time $T_{i_j}$ with $j \in [0,r_{par}-1]$ (parallel sub-algorithms of  $A$).
\end{itemize}
\end{rem}

 \noindent From the sequence   $i=0,\ldots,r_E-1$, numbering the $r_E$ rows of $M_{E_{k,P}}$,   two subsequences of indices originate $\{i_q\}_{q\in [0,r_{seq}-1]}$, and $\{i_r\}_{r\in [0,r_{par}-1]}$,
 and the following definition follows
    \begin{defn}(\textbf{Parallel Execution time})
The quantity
	\begin{equation}\label{TempoPar}
	 T_{par}(A_{k,P}):=\sum_{r=0}^{r_{par}-1} T_{i_r}(A_{k,P})
	\end{equation}
is said  parallel execution time of  $A_{k,P}$.
    \end{defn}

    \begin{defn}(\textbf{Sequential Execution time})
	The quantity
	\begin{equation}\label{TempoSeq}
	 T_{seq}(A_{k,P}):=\sum_{q=0}^{r_{seq}-1} T_{i_q}(A_{k,P})
	\end{equation}
 is  said sequential execution time of  $A_{k,P}$.
    \end{defn}
%\end{minipage}}

\noindent The (\ref{tesec1}) can be written as  %evidenziando la componente del tempo dovuta all'esecuzione sequenziale, e quella all'esecuzione parallela:
\begin{equation}\label{tesec3}
	T(A_{k,P})=T_{seq}(A_{k,P})+T_{par}(A_{k,P})\quad .
\end{equation}
This states that, by looking at matrix $M_{E_{k,P}}$, the model expresses the size of  the parallel and the sequential parts composing the  execution time $A_{k,P}$.\\

Let
  \begin{equation}\label{Rcalc}
  R^{calc}(A_{k,P}):=\frac{\beta_{M_{E_{k,P}}}^{calc}}{r_E} \quad .
  \end{equation}
  $R^{calc}$ is  the parameter of the algorithm $A_{k,P}$  depending  on the most computationally intensive sub-algorithms of $A$.

\noindent    It holds
	\begin{equation}\label{tesec2}
	  T(A_{k,P})= R^{calc}(A_{k,P})\cdot r_E \cdot tcalc=\sum_{r=0}^{r_E-1}\beta_{r,M_{E_{k,P}}}^{calc}\cdot tcalc\quad
	\end{equation}

\begin{rem}\label{tseqR}
    If $P=1$, since $r_E=C(A_{k,1})=k$ from(\ref{Rcalc}) it is
    \begin{equation}\label{27bis}
    R^{calc}(A_{k,1}):=\frac{\beta_{all,M_{E_{k,P}}}^{calc}}{k}.
	\end{equation}
\end{rem}

\begin{cor}\label{cortemposeq}
From the (\ref{Rcalc}) it follows
\begin{equation}\label{coro24}
    T(A_{k,1}) = k\cdot R^{calc}(A_{k,1}) \cdot tcalc \quad ;
	\end{equation}

%\proof
%If  $A_{k,P}$ is  sequential, it means that $c_E=1$ and $r_E=C(A_{k,1})=k$ and from (\ref{tesec2}) thesis follows.
%\endproo
 \begin{equation}\label{tesecP2}
	  T(A_{k,P})\geq r_D\cdot R^{calc}(A_{k,P}) \, tcalc \quad,
\end{equation}
and it assumes its minimum value when $r_E = r_D$.
\begin{equation}\label{thTempoPS}
	  T(A_{k,P})=(r_{seq}+r_{par})\cdot R^{calc}(A_{k,P}) \cdot tcalc \quad .
	\end{equation}
\end{cor}

\begin{defn}(\textbf{Speed up in $\frac{AL}{\rho}$ })\label{sp_gen}
   Given $\mathcal{B}_{N_r}$, two different decompositions $D_{k}(\mathcal{B}_{N_r})$ and $D_{k'}(\mathcal{B}_{N_r})$, and
   \begin{itemize}
      \item $A_{k,P}\in \varphi^{-1}(D_{k}(\mathcal{B}_{N_r}))$, where $P>1$,
      \item $A_{k',1}\in \varphi^{-1}(D_{k'}(\mathcal{B}_{N_r}))$
   \end{itemize}
    where $\mathcal{M}_1$ and $\mathcal{M}_P$ differ only on the number of processing elements, if $\mathcal{G}(A_{k,P})=\mathcal{G}(A_{k',P})$, then the speed up of $A_{k,P}$ with respect to $A_{k',1}$ is
  \begin{equation}\label{sp_general}
   Sp(A_{k,P},A_{k',1}):=Sc_{up}(A_{k,P},A_{k',1})\cdot \frac{T(A_{k,1})}{T(A_{k,P})}=\frac{k'}{k}\cdot \frac{\beta_{sum,M_E(A_{k,P})}^{calc}}{\beta_{M_E(A_{k,P})}^{calc}}\quad .
\end{equation}
\end{defn}

\begin{rem} (Ideal  Speed up)\label{idealspgen}
  Since it is always\footnote{$\beta_{M_E(A_{k,P})}^{calc}$ is the sum of the maximum operator time on each row, so  $\beta_{sum,M_E(A_{k,P})}^{calc}$ can be equal to $P\cdot \beta_{M_E(A_{k,P})}^{calc}$ only if the operators have all the same time.} $$\beta_{sum,M_E(A_{k,P})}^{calc} \leq P\cdot \beta_{M_E(A_{k,P})}^{calc}$$ then it holds that
    \begin{equation}
   Sp(A_{k,P},A_{k',1})\leq Sc_{up}(A_{k,P},A_{k',1})\cdot P\quad .
\end{equation}
\end{rem}

\begin{defn}(\textbf{Speed up in $\rho(A_{k,P})$})\label{sp_classic}
    The speed up of $A_{k,P}$ with respect to $A_{k,1}$ is
  \begin{equation}\label{sp}
   Sp(A_{k,P})=\frac{T(A_{k,1})}{T(A_{k,P})}=\frac{\beta_{sum,M_E(A_{k,P})}^{calc}}{\beta_{M_E(A_{k,P})}^{calc}}\quad .
\end{equation}
\end{defn}

\section{Algorithms which are in the same equivalence class}\label{secPer}
We consider algorithms that are in the same equivalence class, i.e. those corresponding to the same
decomposition of the problem

% \fbox{\begin{minipage}{\textwidth}
    \begin{thm}\label{thmProbPar3}
	$\forall \, \mathcal{B}_{N_r}$ perfectly decomposed according to the decomposition $D_{k}(\mathcal{B}_{N_r})$,
   and   $\forall \, A_{k,P}$ perfectly parallel algorithm that solves it
   on  $\mathcal{M}_P$ with  $P>1$, if $$Cop_{\mathcal{M}_1}\equiv Cop_{\mathcal{M}_p}, \quad $$it follows that:
  \begin{equation}
  \begin{split}
  T(A_{k,P})=\frac{T(A_{k,1})}{P}\cdot \frac{R^{calc}(A_{k,P})}{R^{calc}(A_{k,1})} \quad .
  \end{split}
  \end{equation}
  \proof
  If $A_{k,P}$ is perfectly parallel, then $M_{E_{k,P}}$ has no  empty elements so
  $$r_E=\frac{k}{c_E}=\frac{k}{P}\quad .$$
  Therefore, from the  (\ref{tesec2}) and  \ref{coro24}, it is
  \begin{equation}\label{tesecP}
\begin{split}	
T(A_{k,P})&=r_E\cdot R^{calc}(A_{k,P})\cdot  tcalc\\
&=\frac{k}{c_E}\cdot R^{calc}(A_{k,P}) \cdot tcalc= \frac{T(A_{k,1})}{P}\cdot \frac{R^{calc}(A_{k,P})}{R^{calc}(A_{k,1})} \quad .
  \end{split}
  \end{equation}
  \endproof
    \end{thm}
%\end{minipage}}

\begin{thm}
	For all the matrices  $M_{E_{k,P}}$ of algorithms in $\varrho(A_{k,P})$,  it holds
	\begin{equation}\label{ossE}
	 c_E \leq c_{D_k}
	\end{equation}
	and
	\begin{equation}\label{rerd}
	 r_E \geq r_{D_k} .
	\end{equation}
	Moreover, let us consider  $A_{k,P}^i$ and $A_{k,P}^j$ two algorithms belonging to $\varrho(A_{k,P})$, and their matrices $M_{E_{k,P}}^i$ and  $M_{E_{k,P}}^j$. We have:
	\begin{itemize}
	 \item $c_E^i < c_E^j \Rightarrow r_E^i\geq r_E^j$;
	 \item $c_E^i > c_E^j \Rightarrow r_E^i\leq r_E^j$.
	\end{itemize}
	\proof
From inheritance on $A_{k,P}$ of dependencies defined on $D_{k}(\mathcal{B}_{N_r})$,
it is not possible
that $c_E > c_D$, therefore $c_E\leq c_{D_k}$. Then there is at least one row of
$M_{D_k}$ with $c_{D_k}$ non-empty elements.
Let $d$ be the difference between $c_{D_k}$ and $c_E$. Therefore, since $M_{D_k}$ and
$M_E$ have the same number of non-empty elements, it is $r_E\ge r_D +\lceil(d/c_E.)\rceil$.

Similarly, it can
be proved that if $c^i_E < c^j_E$ then $r^i_E \ge r^j_E$, and  if $c^i_E > c^j_E$ then $r^i_E \le r^j_E$.
	\endproof
\end{thm}
%\end{minipage}}

\begin{rem} The minimum execution time is proportional to the dependency degree of $\mathcal{B}_{N_r}$, that is when the number of computing units is equal to the  concurrency degree  of $\mathcal{B}_{N_r}$.
 \end{rem}

We now  define a subset of the equivalence class of $\varrho(A_{k,P})$. Let $\simeq$ be the equivalence relation identifying two algorithms with the same $P$. Then
\begin{equation}\label{hatrho}
  \hat \varrho(A_{k,P}):=  \varrho(A_{k,P})/\simeq
\end{equation}
i.e. consisting of the representatives of the equivalence classes of $\simeq$\footnote{For example, we can take  the algorithm in $\hat \varrho(A_{k,P})$,  $P\geq 1$, whose execution matrix has the fewest number of  rows.}.

Let us now consider matrices $M_{E_{k,P}}$ associated to algorithms belonging to  $\hat \varrho(A_{k,P})$, varying $P$.

The following result defines the speed up of a parallel algorithm with respect to the sequential algorithm belonging to its class.

\begin{thm}
  Consider$A_{k,1}\buildrel \varrho \over \equiv A_{k,P}$ with
	  $$M_{E_1}\text{, of order } N_{1}^{E}=r_{E_1}\cdot 1 \text{ and } M_{E_P}\text{, of order } N_{P}^{E}=r_{E_P}\cdot P.$$ It holds
  \begin{equation}\label{sp2}
    Sp(A_{k,P})=\frac{\beta_{sum,M_{E_{k,1}}}^{calc}}{r_{E_P}\cdot R^{calc}(A_{k,P})}\quad .
    \end{equation}
    \proof
	  From the (\ref{tesec2}),  (\ref{27bis}) and (\ref{sp}), it follows
	  \begin{equation}\label{speq}
	  	\begin{split}	
 Sp(A_{k,P})&=\frac{r_{E_1}\cdot R^{calc}(A_{k,1}) \cdot  tcalc}{r_{E_P}\cdot R^{calc}(A_{k,P})  \cdot tcalc}=
 \frac{C(A_{k,P})}{r_{E_P}} \frac{R^{calc}(A_{k,1})}{R^{calc}(A_{k,P})}\\
 &=\frac{\beta_{sum,M_{E_{k,1}}}^{calc}}{r_{E_P}\cdot R^{calc}(A_{k,P})}.
		\end{split}
    \end{equation}
    \endproof
\end{thm}

\begin{cor}\label{idealsp}
    Since $(r_{E_P}\cdot c_{E_P})\geq C(A_{k,P})$,  from the (\ref{speq}) it follows  that
    $$Sp(A_{k,P})\leq c_{E_P}\cdot \frac{R^{calc}(A_{k,1})}{R^{calc}(A_{k,P})}=P\cdot \frac{R^{calc}(A_{k,1})}{R^{calc}(A_{k,P})}\quad .$$
\end{cor}

\begin{defn}\textbf{(Ideal Speed up in $\hat \varrho(A_{k,P})$)}
    We let
    \begin{equation}\label{idealsp}
      Sp_{Ideal}(A_{k,P}) =%\max_{\hat \varrho(A_{k,P})} Sp(A_{k,P})=
    P\cdot \frac{R^{calc}(A_{k,1})}{R^{calc}(A_{k,P})}\quad ,
    \end{equation}
    be the ideal speed up.
\end{defn}

Let  $r_{par_i}$  denote the number of rows  having  $i> 1$ not empty elements, and $r_{par_1}=r_{seq}$,  then it is  $$r_{E_P}=\sum_{i=1}^{P} r_{par_i}\quad .$$

\begin{defn}(\textbf{Total  Time of $A$ with $i$ non empty elements})
Let $T_{j_i}$ the time of a row with  $i\geq 1$  not empty elements elements.
The quantity
	\begin{equation}\label{TempoParI}
	 T_{par_i}(A_{k,P})=\sum_{j=0}^{r_{par_i}-1} T_{i_j}
	\end{equation}
is the execution time   of the part of $A$ with  $i$ non empty elements on each row.
    \end{defn}

    \begin{rem}
    It holds that

  $r_{par}=r_{E_P}-r_{seq}=\sum_{i=2}^{P} r_{par_i}$  then  $T_{par_1}(A_{k,P})=T_{seq}(A_{k,P}).$

\end{rem}

 Next result  shows how the generalized Amdhal's Law can be derived  by using the rows of the execution matrix $M_{E_{k,P}}$ having at least one non empty element.

     \begin{thm}(\textbf{Generalized Amdhal's Law})
    It is
    \begin{equation}\label{WareG}
    Sp(A_{k,P})=\frac{\frac{R^{calc}(A_{k,1})}{R^{calc}(A_{k,P})}}{\sum_{i=1}^{P} \alpha_i}\quad ,
    \end{equation}
    where
    $$\alpha_i=\frac{r_{par_i}}{C(A_{k,P})}\quad .$$
   \begin{proof}
    From  (\ref{speq}) it is
     \begin{equation}
	  Sp(A_{k,P})=\frac{\frac{C(A_{k,P})\cdot R^{calc}(A_{k,1})}{R^{calc}(A_{k,P})}}{r_{seq}+\sum_{i=2}^{P} r_{par_i}}\quad .
    \end{equation}
    By dividing for  $C(A_{k,P})$ it follows that
    \begin{equation}
	  Sp(A_{k,P})=\frac{\frac{R^{calc}(A_{k,1})}{R^{calc}(A_{k,P})}}{\frac{r_{seq}}{C(A_{k,P})}+\sum_{i=2}^{P} \frac{r_{par_i}}{C(A_{k,P})}}
    \end{equation}
    that is
    \begin{equation} Sp(A_{k,P})=\frac{\frac{R^{calc}(A_{k,1})}{R^{calc}(A_{k,P})}}{\alpha_1+\sum_{i=2}^{P} \alpha_i}\quad . \end{equation}
   \end{proof}
    \end{thm}

Then, the Amdhal's Law \cite{amdahl} comes out as a particular case of the previous theorem
    \begin{cor}\label{Ware}(\textbf{Amdhal's Law})
    If we assume that  $M_{E_{k,1}}$ only has rows with $1$ element or  $P$ elements, we have
    \begin{equation}\label{48}
    Sp(A_{k,P})=\frac{\frac{R^{calc}(A_{k,1})}{R^{calc}(A_{k,P})}}{\alpha + \frac{1-\alpha}{P}}\quad ,
    \end{equation}
    where
    $$\alpha:=\frac{r_{seq}}{C(A_{k,P})}\quad .$$
   \proof
    From (\ref{WareG}) it follows that
     \begin{equation}
	  Sp(A_{k,P})=\frac{\frac{R^{calc}(A_{k,1})}{R^{calc}(A_{k,P})}}{\alpha_1+\sum_{i=2}^{P} \alpha_i}\quad ,
    \end{equation}
    where
    $$\alpha_i:=\frac{r_{par_i}}{C(A_{k,P})}$$
    and
    $$\frac{r_{par}}{C(A_{k,P})}=\sum_{i=2}^{P} \alpha_i \quad .$$
    If the rows with more than one non empty element have $P$ elements, it is
    $$r_{par}=\frac{C(A_{k,P})-r_{seq}}{P}$$
    therefore, if we let $\alpha_1=\alpha=\frac{r_{seq}}{C(A_{k,P})}$ we  get
    \begin{equation}
    \begin{aligned} Sp(A_{k,P})&=\frac{\frac{R^{calc}(A_{k,1})}{R^{calc}(A_{k,P})}}{\frac{r_{seq}}{C(A)}+\frac{r_{par}}{C(A_{k,P})}}=\frac{\frac{R^{calc}(A_{k,1})}{R^{calc}(A_{k,P})}}{\frac{r_{seq}}{C(A_{k,P})}+\frac{C(A_{k,P})-r_{seq}}{C(A_{k,P})\cdot P}}&=\frac{\frac{R^{calc}(A_{k,1})}{R^{calc}(A_{k,P})}}{\alpha+\frac{1-\alpha}{P}}.
    \end{aligned}
    \end{equation}
   \endproof
    \end{cor}

\noindent Let  $Q$ denote the cost of  $A_{k,P}$. The cost is defined as the product of  the execution time and the number of processors utilized \cite{kennedy}. In this mathematical settings it holds that
the cost $Q$ can be written as
  \begin{equation}\label{costo2}
	   Q(A_{k,P})=c_{E}\cdot r_{E}\cdot R^{calc}(A_{k,P}) \cdot  tcalc\quad .
    \end{equation}
%\end{minipage}}

\noindent  If $c_E=1$, from the (\ref{coro24}) it holds
  \begin{equation}
  \begin{split}
	  Q(A_{k,1})&=r_{E}\cdot R^{calc}(A_{k,P}) \cdot tcalc=T(A_{k,1})=C(A_{k,P})\cdot R^{calc}(A_{k,1})\cdot tcalc\\
	  &=\beta_{sum,M_{E_{k,1}}}^{calc} \cdot tcalc\quad .
	  \end{split}
  \end{equation}

The overhead of $A_{k,P}$ is the total time spent by all the processing elements over and above that spent in useful computation.

\begin{defn}\label{overh}(\textbf{Algorithm Overhead})
	The quantity
	\begin{equation}\label{overeq}
		Oh(A_{k,P}):=Q(A_{k,P})-Q(A_{k,1})=\left(c_E\cdot \beta_{M_{E_{k,P}}}^{calc}-\beta_{sum,M_{E_{k,1}}}^{calc}\right)\cdot tcalc \quad .
	\end{equation}
is said overhead of  $A_{k,P}$.
\end{defn}
%\end{minipage}}\vspace{0.1cm}

%\fbox{\begin{minipage}{\textwidth}
\begin{thm}\label{thOverh0}
	It holds
\begin{equation}\label{alovh}
  C(A_{k,P})\cdot (R^{calc}(A_{k,P})- R^{calc}(A_{k,1}))\cdot tcalc  \left \{
  \begin{array}{cc}
    =0 & \text{ if } R^{calc}(A_{k,P})=R^{calc}(A_{k,1}) \\
    >0 & \text{ otherwise. }
  \end{array}
\right .
\end{equation}
\proof
It holds
\begin{equation}
\begin{split}
   Q(A_{k,P})&\geq card(A_{k,P})\cdot R^{calc}(A_{k,P})\cdot tcalc\\
   &=C(A_{k,P})\cdot R^{calc} (A_{k,P})\cdot tcalc=k\cdot R^{calc} (A_{k,P})\cdot tcalc
\end{split}
\end{equation}
Moreover,
$$Q(A_{k,1})=C(A_{k,1})\cdot R^{calc}(A_{k,1})\cdot tcalc=k\cdot R^{calc}(A_{k,1})\cdot tcalc$$
therefore it follows from (\ref{overeq})
$$Oh(A_{k,P})\geq \left(k\cdot (R^{calc}(A_{k,P})- R^{calc}(A_{k,1}))\right)\cdot tcalc$$
and the (\ref{alovh}) follows.
\endproof
\end{thm}

\begin{defn}\label{ovIdeal}
\textbf{(Ideal Overhead in $\hat \varrho(A_{k,P})$)} \\
    From the (\ref{alovh}) it follows
    \begin{equation}\label{idealovh}
    \begin{split}
        Oh_{Ideal}(A_{k,P})=%\max_{\hat \varrho(A_{k,P})} Oh(A_{k,P})\\
        \left(k\cdot (R^{calc}(A_{k,P})- R^{calc}(A_{k,1})\right)\cdot tcalc\quad .
    \end{split}
    \end{equation}
\end{defn}

Let  $Ef(A_{k,P}):=\frac{Sp(A_{k,P})}{P}$ be the efficiency of $A$ where $P\geq 1$.
\begin{thm}\label{thEf}
  Let $N^E_P=c_{E_P}\cdot r_{E_P}$, denote the dimension of the execution matrix of $A_{k,P}$, it holds that
    \begin{equation}\label{ef2}
	  Ef(A_{k,P})=\frac{\beta_{sum,M_{E_{k,1}}}^{calc}}{N^E_P\cdot R^{calc}(A_{k,P})}\quad .
    \end{equation}
\proof
    Since $c_{E}=P$, it follows that
    \begin{equation}
	Ef(A_{k,P})=\frac{Sp(A_{k,P})}{P}=\frac{\beta_{sum,M_{E_{k,1}}}^{calc}}{c_{E_P}\cdot r_{E_P}\cdot R^{calc}(A_{k,P})}\quad .
	    \end{equation}
    \endproof
\end{thm}

\begin{defn} \label{idealeff}
\textbf{(Ideal Efficiency in $\hat \varrho(A_{k,P})$)} \\
Since $Sp(_{k,P})\leq P\cdot \frac{R^{calc}(A_{k,1})}{R^{calc}(A_{k,P})}$,  it  always is $Ef(A_{k,P})\leq \frac{R^{calc}(A_{k,1})}{R^{calc}(A_{k,P})}$. So let
\begin{equation}\label{idealefficiency}
  Ef_{Ideal}(A_{k,P})=%\min_{\hat \varrho(A_{k,P})} Ef(A_{k,P})=
\frac{R^{calc}(A_{k,1})}{R^{calc}(A_{k,P})}\quad .
\end{equation}

be the ideal efficiency of $A_{k,P}$.
\end{defn}

\begin{rem} It is worth to note the role of  parameters $R^{calc}(A_{k,P})$ and $R^{calc}(A_{k,1})$ in  (\ref{48}), (\ref{idealovh}) and  (\ref{ef2}). If in $A_{k,P}$ there are few operators which are much more time consuming than the others, and $k>>r_E$ then
$\beta_{M_{E_{k,P}}}^{calc} \simeq \beta_{sum,M_{E_{k,1}}}^{calc}$ and $R^{calc}(A_{k,P}) >> R^{calc}(A_{k,1})$.   The more the operators  are and the greater the difference is in  (\ref{idealovh}), or the lower the ratio is in (\ref{48}) and  (\ref{ef2}). Hence, the greater the overhead is, the lower the speed up and the efficiency are.  This is  a consequence of a  problem decomposition, associated to $A_{k,P}$  not well balanced.\\
\end{rem}

Let us now suppose that the algorithm $A_{k,P}$ is perfectly parallel, that is its execution matrix $M_{E_P}$ has not any empty element.
Since $r_{E_P}\cdot c_{E_P}=C(A_{k,P})$ it follows from  (\ref{idealsp}) that
$$Sp(A_{k,P})=Sp_{Ideal}(A_{k,P})=P\cdot \frac{R^{calc}(A_{k,1})}{R^{calc}(A_{k,P})}\quad ,$$
from  (\ref{alovh}) that
$$Oh(A_{k,P})=Oh_{Ideal}(A_{k,P})=\left(C(A_{k,P})\cdot (R^{calc}(A_{k,P})- R^{calc}(A_{k,1}))\right)\cdot tcalc\quad ,$$
from (\ref{idealefficiency})
$$Ef(A_{k,P})=Ef_{Ideal}(A_{k,P})=\frac{R^{calc}(A_{k,1})}{R^{calc}(A_{k,P})}\quad .$$

%\fbox{\begin{minipage}{\textwidth}
\begin{rem}
   If $P=c_D$, $r_E=r_D$ and $c_E=c_D$, if $P=c_D$ then  the following results hold on:
   \begin{enumerate}
     \item $Q(A_{k,P})=c_D\cdot r_D\cdot  R^{calc}(A_{k,P}) \cdot tcalc= N_{D}\cdot R^{calc}(A_{k,P}) \cdot tcalc $;
     \item $Sp(A_{k,P})=\frac{C(A_{k,P})}{r_D}\frac{R^{calc}(A_{k,1})}{R^{calc}(A_{k,P})};$
     \item $Oh(A_{k,P})=(c_D \cdot r_D - C(A_{k,P}))\cdot  R^{calc}(A_{k,P}) \cdot tcalc;$
     \item $Ef(A_{k,P})=\frac{C(A_{k,P})}{r_D\cdot c_D}\frac{R^{calc}(A_{k,1})}{R^{calc}(A_{k,P})}\quad .$
   \end{enumerate}
 \end{rem}

\section{Algorithms with operators having the same execution time}\label{secPer2}
We assume that  all the operators of the algorithm have the same execution time. For example they are the elementary floating point operations. The execution time is  $\beta^{calc} \cdot tcalc$, and without loss of generality we assume that $\beta^{calc}=1$.
Hence, it follows that, $$\forall P, \quad \beta_{r,M_{E_{k,P}}}^{calc}=1, \quad \beta_{M_{E_{k,P}}}^{calc}=r_E, \quad \beta_{sum,M_{E_{k,P}}}^{calc}=k\quad .$$ Finally, from  (\ref{Rcalc}) it follows that $$\forall P, \quad R^{calc}(A_{k,P})=1\quad .$$
Hence, we get
\begin{itemize}
\item  $Sp(A_{k,P},A_{k',1}):=\frac{k'}{k}\cdot \frac{k}{r_{E}}\quad ,$
\item if $Q=1$, then  $Sp(A_{k,P}):=\frac{k}{r_E}\quad ,$
\item  $Sp_{Ideal}(A_{k,P},A_{k',1}) = Sc_{up}(A_{k,P},A_{k',1})\cdot P=\frac{k'}{k}\cdot P\quad ,$
\item $Sp_{Ideal}(A_{k,P}) = c_{E_P}=P\quad .$
\end{itemize}
 Finally, if  $\mathcal{B}_{N_r}$ is perfectly decomposed then
  \begin{equation}
  T(A_{k,P})=\frac{T(A_{k,1})}{P}\quad ,
  \end{equation}
 i.e.  $A_{k,P}$ has the ideal speed up in the classical definition.\\

Let us now consider matrices $M_{E_{k,P}}$ associated with algorithms in  $\hat \varrho(A_{k,P})$, varying $P$. The following results hold
\begin{enumerate}
  \item $Q(A_{k,P})=c_{E}\cdot r_{E}\cdot  tcalc $ and, if $c_E=1$, then $Q(A_{k,1})=k \cdot tcalc\quad ;$
  \item $Oh_{Ideal}(A_{k,P})=0\quad ;$
  \item $Ef_{Ideal}(A_{k,P}) = 1\quad .$
\end{enumerate}

Finally, next result relates the overhead to the sparsity degree of the execution matrix.
% \fbox{\begin{minipage}{\textwidth}
\begin{thm}
Let suppose that
	\begin{equation}\label{tcalcall}
	    \forall I^{i_j}\in Cop_{\mathcal{M}_P}, \quad t_{i_j}=\beta^{calc}_{i_j,M_{E_{k,1}}} \cdot tcalc=tcalc, \quad \forall i,j\quad .
	\end{equation}
 Given
  $A_{k,P}$, $ P>1$,  $M_{E_{k,P}}$ of order $N^{E}_P=r_{E}\cdot P$,
 let $V_r$ be the number of empty elements of the row $r$ of  $M_{E_{k,P}}$; it is
  \begin{equation}\label{overh3}
	  Oh(A_{k,P})= \sum_{r=0}^{r_E-1} V_r\cdot tcalc \quad .
  \end{equation}
  \begin{proof}
 It holds that
  $$c_E \cdot r_E=card(A_{k,P})+\sum_{r=0}^{r_{E}-1} V_r=C(A_{k,P})+\sum_{r=0}^{r_{E}-1} V_r=k+\sum_{r=0}^{r_{E}-1} V_r$$
	then from (\ref{overeq})
	\begin{equation}
		Oh(A_{k,P})= \left(k+\sum_{r=0}^{r_{E}-1} V_r - k\right)\cdot tcalc=\sum_{r=0}^{r_E-1} V_r \cdot tcalc \quad .
	\end{equation}
\end{proof}
\end{thm}
\begin{rem}
   Note that $\sum_{r=0}^{r_E-1} V_r$ is the sparsity degree of the execution matrix.
\end{rem}

\noindent Following table collects the expressions of the quantities that we have derived and that characterize the mathematical framework.

%\newpage

\normalsize

%The number and size of sub-problems  into which a problem is decomposed determine the granularity of the decomposition. A decomposition into a large number of small computational problems is called fine-grained and a decomposition into a small number of large computational problems  is called coarse-grained.
 %Here, granularity has a major consequence in the level of detail that is required for an algorithm to be analysed with this approach.
 Among the decomposition approaches,  recursive decomposition is the most suitable for our performance model, especially for a real-world algorithm. In this case, as described in the example below, a problem is solved by first decomposing  it into a set of independent sub-problems. Furthermore, each one of these sub-problems is solved by  applying a similar decomposition into smaller subproblems followed by a combination of their results, and so on. % (an example of this recursive application of the decomposition is in \cite{keyPPAM}, where a first step in the development of the present model is described.).
 In this way we get a decomposition matrix whose elements  can be subsequently decomposed until the desired level of detail  which is considered the most suitable for the subsequent analysis. \\

\footnotesize{
\noindent \emph{Example}: Let $\mathcal{B}_{16}$ denote the computational problem of the sum of $16$ real numbers and $D_{3}(\mathcal{B}_{16})=\{B_8,B_8,B_2\}\in \mathcal{D}\mathcal{B}_{16}.$ The decomposition matrix is
\begin{equation}
  M_{D_3}(\mathcal{B}_{16})=
  \begin{bmatrix}
  \mathcal{B}_8 & \mathcal{B}_8\\
  \mathcal{B}_2 & \varnothing \\
  \end{bmatrix}.
\end{equation}
If $\mathcal{B}_8$ can be decomposed as $D^1_3(\mathcal{B}_8)=\{\mathcal{B}_4,\mathcal{B}_4,\mathcal{B}_2\}\in \mathcal{D}\mathcal{B}_{8}$ then
\begin{equation}
  M_{D^1_3}(\mathcal{B}_{8})=
  \begin{bmatrix}
  \mathcal{B}_4 & \mathcal{B}_4\\
  \mathcal{B}_2 & \varnothing  \\
  \end{bmatrix}.
\end{equation}
In the same way, if $\mathcal{B}_4$ can   be decomposed as
$D^2_3(\mathcal{B}_{4})=\{\mathcal{B}_2,\mathcal{B}_2,\mathcal{B}_2\}\in \mathcal{D}\mathcal{B}_{8} $
and
\begin{equation}
  M_{D^2_3}(\mathcal{B}_{4})=
  \begin{bmatrix}
  \mathcal{B}_2 & \mathcal{B}_2\\
  \mathcal{B}_2 & \varnothing \\
  \end{bmatrix}.
\end{equation}
We have three decompositions for $\mathcal{B}_{16}$:
\begin{equation}
 \begin{aligned}
  D_{3}\in \mathcal{D}(\mathcal{B}_{16})&= \{\mathcal{B}_8,\mathcal{B}_8,\mathcal{B}_2\}.\\
  D_{7}\in \mathcal{D}(\mathcal{B}_{16})&\equiv D_3^1(\mathcal{B}_8)\cup D_3^1(\mathcal{B}_8) \cup \{\mathcal{B}_2\}\\
			      &\equiv \{\mathcal{B}_4,\mathcal{B}_4,\mathcal{B}_2\}\cup \{\mathcal{B}_4,\mathcal{B}_4,\mathcal{B}_2\}\cup \{\mathcal{B}_2\}.\\
  D_{15}\in \mathcal{D}(\mathcal{B}_{16})&\equiv D_3^2(\mathcal{B}_4)\cup D_3^2(\mathcal{B}_4) \cup \{\mathcal{B}_2\}\cup D_3^2(\mathcal{B}_4)\cup D_3^2(\mathcal{B}_4) \cup \{\mathcal{B}_2\}\cup \{\mathcal{B}_2\}\\
			      &\equiv \{\mathcal{B}_2,\mathcal{B}_2,\mathcal{B}_2\}\cup \{\mathcal{B}_2,\mathcal{B}_2,\mathcal{B}_2\}\cup \{\mathcal{B}_2\}\cup \{\mathcal{B}_2,\mathcal{B}_2,\mathcal{B}_2\}\cup \{\mathcal{B}_2,\mathcal{B}_2,\mathcal{B}_2\}\cup \{\mathcal{B}_2\}\cup \{\mathcal{B}_2\}\\
			      &\equiv \{\mathcal{B}_2^i\}_{0\leq i< 15}\in \mathcal{D}\mathcal{B}_{16}.
 \end{aligned}
\end{equation}
%di cui le matrici di dipendenza sono rappresentate nella figura \ref{fig:divImpfigures} e
with the following characteristics, according to the corresponding decomposition matrices:
\begin{itemize}
 \item $D_{3}$: cardinality $3$, concurrency degree $2$  and dependence degree $2$,
 \item $D_{7}$: cardinality $7$, concurrency degree $4$ and dependence degree $3$,
 \item $D_{15}$: cardinality $15$, concurrency degree $8$ and dependence degree $4$.
\end{itemize}
meaning that the intrinsic concurrency of a problem heavily depends on the decomposition chosen for that problem. Each decomposition
has a level of detail depending on the type of subproblems that are  considered. }\\[.1cm]

\noindent
\footnotesize
\begin{center}
%{\large The mathematical framework}\\[.2cm]
\begin{tabular}{||l||}
  \hline
  \\
  % after \\: \hline or \cline{col1-col2} \cline{col3-col4} ...
  Algorithm's performance metrics
  $A_{k,P} \equiv A_{D_{k}(\mathcal{B}_{N_r}),\mathcal{M}_{P}}\in \varrho(A) $\\[0.3cm] \hline \hline
  \\
  $C(A_{k,P}):= k$ $\forall P$ \\[0.2cm]
  $T_r(A_{k,P}):=\beta_{r,M_{E_{k,P}}}^{calc} \cdot tcalc$  \\[.2cm]
  $T(A_{k,P}):=\beta_{M_{E_{k,P}}}^{calc}\cdot tcalc$  \\[0.2cm]
  $Sc_{up}(A_{j,P}, A_{i,P}):=\frac{j}{i}$, $i\leq j$\\[.2cm]
  $Sp(A_{k,P},A_{k',Q}):=\frac{k'}{k}\cdot \frac{\beta_{M_{E_{k',Q}}}^{calc}}{\beta_{M_{E_{k,P}}}^{calc}}     $, $k\leq k'$, $1\leq Q <P$ \\[0.2cm]
  $Sp(A_{k,P},A_{k,Q}):= \frac{\beta_{M_{E_{k',Q}}}^{calc}}{\beta_{M_{E_{k,P}}}^{calc}}$ $1\leq Q <P$, \\[0.2cm]
  $Sp(A_{k,P}):= \frac{\beta_{all,M_{E_{k,P}}}^{calc}}{\beta_{M_{E_{k,P}}}^{calc}}$ \\[0.2cm]
  $Q(A_{k,P}):=c_E\cdot r_E\cdot R^{calc}(A_{k,P})tcalc$  \\[0.2cm]
  $Oh(A_{k,P}):=\left(c_E\cdot \beta_{M_{E_{k,1}}}^{calc}-\beta_{all,M_{E_{k,1}}}^{calc}\right)\cdot tcalc$ \\[0.2cm]
  $Ef(A_{k,P}):=\frac{\beta_{sum,M_{E_{k,P}}}^{calc}}{r_E\cdot c_E}\frac{R^{calc}(A_{k,1})}{R^{calc}(A_{k,P})}$ \\[0.2cm]
  %$P_{op}(A_{k,P}):= argmax_{P\in [a,b]} \frac{C(A_{k,P})}{r_E^2\cdot c_{E}\cdot R^{calc}(A_{k,P}) \cdot tcalc}$\\[0.3cm]
  \hline \hline
  \\
  Algorithm's parameters \\[.2cm]\hline \hline
  \\
   $\beta_{r,M_{E_{k,P}}}^{calc}:=\max_{j\in [0,c_E-1]} \beta^{calc}_{r_j,M_{E_{k,P}}} $ \\[.2cm]
   $\beta_{M_{E_{k,P}}}^{calc}:=\sum_{r=0}^{r_E-1} \beta_{r,M_{E_{k,P}}}^{calc}$ \\[.1cm]
  $R^{calc}(A_{k,P}):=\frac{\beta_{M_{E_{k,P}}}^{calc}}{r_E}\geq 1$\\[.2cm]
\hline \hline
\\
  Machine's parameters  of $\mathcal{M}_P:= Cop_{\mathcal{M}_P}$\\[.2cm]\hline \hline
  \\
  $t_{i_j}:=\beta^{calc}_{i_j,M_{E_{k,P}}} \cdot tcalc $ \\[.2cm]
\hline \hline
\end{tabular}
\end{center}

\normalsize

\normalsize

\section{Conclusion}\label{secConc}
Recent activities of major chip manufacturers show more evidence
than ever that future designs of microprocessors and large  systems will be heterogeneous in nature,
relying on the integration of two major types of components.   On the first hand,
multi/many-cores CPU technology  have  been  developed  and  the  number  of  cores  will  continue  to  escalate  because  of the need to pack more and more components on a chip. On the other hand special purpose hardware and accelerators, especially Graphics Processing Units  are in commodity production. Finally, reconfigurable architectures such as Field Programmable Gate Arrays  offer several parameters such as operating frequency, precision, amount of memory, number of computation units, etc. These parameters define a large design space that must be explored to find  efficient solutions \cite{kuon}. To cope with  this scenario,  performance analysis of parallel algorithms should be  re-evaluated to find out the best-practice algorithm on novel architectures \cite{demmel1,demmel13,gunnels01a,gunnels01b,miller14,voevodin}. In this paper we presented a mathematical framework which can be used to get a multilevel description of a parallel algorithm, and we proved that it can be  suitable for analysing the mapping of the algorithm on a given machine. The model allows the choice of a level of abstraction of the problem decomposition and of the algorithm determining the level of granularity of the performance analysis. This feature can be very useful for analysing the mapping of the algorithm on novel architectures. 
We have  assumed abstract models for both the algorithms and the architectures and made numerous simplifying assumptions. However, we believe that a simplified  parameterized  model gives an useful generalization for better understanding algorithms that can run really fast no matter how complicated the underlying computer architecture \cite{DemmelLesson}.

\end{document}